\documentclass[journal]{IEEEtran}

\usepackage{booktabs}
\usepackage{multirow}
\usepackage{graphicx}
\usepackage{amsmath}
\usepackage{amssymb}
\usepackage{xcolor}
\usepackage{pgfplots}
\usepackage{pgfplotstable}
\usepackage{algorithm}
\usepackage{algpseudocode}
\usepackage{hyperref}
\pgfplotsset{compat=1.18}
\usetikzlibrary{positioning,shapes.geometric,arrows.meta}

\title{Agentic Verifier-in-the-Loop Solver Orchestration for Cell-Free Massive MIMO Downlink Power Control}
\author{Zhichao~Gao,~\IEEEmembership{Graduate Student Member,~IEEE}%
\thanks{E-mail: sdpygzc@163.com. ORCID: \href{https://orcid.org/0009-0006-4579-8433}{0009-0006-4579-8433}.}}

\begin{document}
\maketitle

\begin{abstract}
Cell-free massive multiple-input multiple-output (MIMO) systems can provide uniformly strong service through distributed access points, but performance still depends critically on downlink power control. Existing methods are typically selected offline and then applied uniformly across channel and load regimes, even though no single solver is uniformly best: strong solvers can be slow, fast solvers can fail near feasibility boundaries, and lightweight heuristics can incur larger quality loss. We therefore propose \textbf{VISO-PC}, a verifier-in-the-loop solver-orchestration framework in which an agent routes among trusted solvers rather than generating power coefficients directly. Given a structured instance descriptor, the router selects an initial solver and fallback order, and an independent verifier accepts only candidates that satisfy the constraints and produce a valid verified common rate. For fairness-oriented downlink cell-free power control under per-AP constraints, verification-aware orchestration improves accepted rate over all fixed single-solver baselines on a reproducible prototype benchmark. Moreover, a lightweight memory-based router matches the accepted rate of a strong rule-based router while reducing average runtime and fallback rate. These results show that solver orchestration is a practical agentic layer for cell-free massive MIMO downlink power control.
\end{abstract}


\section{Introduction}

Cell-free massive MIMO provides user-centric service through distributed access points and depends heavily on downlink power control~\cite{Ngo2015CellFreeUniformlyGreat,Ngo2017CellFreeVersusSmallCells}. Classical studies have developed effective methods for fairness and spectral-efficiency objectives~\cite{Nayebi2017PrecodingPowerOptimization,Chakraborty2019CentralizedDistributedMaxMin,Bjornson2020ScalableCellFree,Bjornson2020MakingCellFreeCompetitive}, but they are typically selected offline and then applied uniformly. In practice, no single solver is uniformly best across heterogeneous regimes.

Recent work has introduced agentic and LLM-based intelligence into wireless systems and O-RAN, including tool-oriented communication frameworks~\cite{liu2026unleashing}, intent-driven cell-free O-RAN optimization~\cite{shokouhi2026agentic}, multi-scale O-RAN agent architectures~\cite{navidan2026toward,elkael2025agentran}, and conflict-aware rApp orchestration~\cite{li2026multiagentic}. Other studies consider direct LLM-based wireless optimization~\cite{zhou2024llmenabled} or use LLMs as optimization assistants and scheduler-generation tools~\cite{peng2025llmoptira,elkael2025allstar}, for example to reformulate non-convex problems or synthesize scheduler logic.

Despite their relevance, these works do not address the problem studied here. The closest scenario-level paper is~\cite{shokouhi2026agentic}, but it emphasizes inter-agent coordination rather than verifier-aware solver routing. Direct LLM optimization papers such as~\cite{zhou2024llmenabled} generate decisions rather than select among trusted solvers. LLM-assisted optimization and scheduler-generation systems~\cite{peng2025llmoptira,elkael2025allstar} target formulation, code synthesis, or policy generation rather than runtime orchestration over a fixed solver portfolio. More broadly, O-RAN agent frameworks~\cite{elkael2025agentran,navidan2026toward,li2026multiagentic} study intent decomposition and cross-application coordination, whereas this correspondence focuses on verifier-aware routing among trusted numerical solvers for cell-free power control.

This distinction matters because a deployment-oriented design should preserve feasibility guarantees and should not rely on an LLM to emit continuous control variables directly. The proposed formulation therefore places the agent one level above the optimizer: it selects among trusted numerical tools, while verification and fallback preserve robustness when the initial solver choice is mismatched to the operating regime.

Accordingly, this paper studies a narrower question: can an agent reliably orchestrate a portfolio of trusted solvers for cell-free downlink power control? The proposed \textbf{VISO-PC} framework summarizes each instance into a structured descriptor, selects an initial solver and fallback order, verifies each returned solution independently, and reroutes when the selected solver fails or underperforms. The focus is on downlink, fixed beamforming, per-AP power constraints, and a fairness-oriented objective family, thereby isolating solver orchestration from broader O-RAN concerns.

The main contributions are threefold. First, we propose a verifier-in-the-loop orchestration framework for cell-free downlink power control. Second, we assemble a solver portfolio with complementary quality-latency trade-offs. Third, we establish an evaluation protocol based on verified feasibility, selection regret, fallback rate, and runtime. On the prototype benchmark, orchestration improves accepted rate over fixed single-solver policies, while a lightweight memory-based router improves efficiency relative to a strong rule-based baseline.

\noindent\textit{Notation:} Lowercase letters denote scalars, lowercase boldface letters denote vectors, and uppercase boldface letters denote matrices. Calligraphic symbols denote sets, for example $\mathcal{T}$ for the solver portfolio and $\mathcal{F}$ for the fallback list. The integers $L$ and $K$ denote the numbers of APs and UEs, respectively. The scalar $\eta_{l,k}$ denotes the power-control coefficient from AP $l$ to UE $k$, the vector $\mathbf{x}$ denotes the instance descriptor used by the orchestration layer, and $\pi(\mathbf{x})$ denotes the solver plan selected for the current instance. The quantity $r_{\mathrm{ver}}$ denotes the verified common rate computed by the independent checker. Finally, $\mathbf{1}[\cdot]$ denotes the indicator function, which equals one when the enclosed condition is true and zero otherwise.

\section{System Model and Problem Formulation}

Consider the downlink of a cell-free massive MIMO network with $L$ distributed APs serving $K$ UEs~\cite{Ngo2015CellFreeUniformlyGreat,Ngo2017CellFreeVersusSmallCells}. In this correspondence, each AP and UE has a single antenna, and the beamforming rule is fixed to maximum-ratio transmission. Let $g_{l,k}$ denote the effective channel coefficient from AP $l$ to UE $k$, and let $\eta_{l,k} \ge 0$ denote the power-control coefficient allocated by AP $l$ to UE $k$. The transmit-power budget of AP $l$ is $P_l^{\max}$, so the per-AP power constraint is
\begin{equation}
\sum_{k=1}^{K} \eta_{l,k} \le P_l^{\max}, \qquad \forall l.
\label{eq:papc}
\end{equation}

Under fixed beamforming, the received downlink signal at UE $k$ is a coherent superposition of the desired terms from serving APs plus inter-user interference and noise. This induces an effective signal-to-interference-plus-noise ratio $\mathrm{SINR}_k(\boldsymbol{\eta})$, from which the achievable spectral efficiency is computed as
\begin{equation}
\mathrm{SE}_k(\boldsymbol{\eta})=\log_2\!\big(1+\mathrm{SINR}_k(\boldsymbol{\eta})\big).
\label{eq:se}
\end{equation}
Although the exact closed-form expression depends on the underlying propagation model and channel-hardening assumptions, the framework only requires the verifier to recompute $\mathrm{SE}_k$ consistently from a returned candidate solution.

The target task is restricted to a fairness-oriented family, represented by the max-min common-rate problem
\begin{equation}
\max_{\boldsymbol{\eta}} \; \min_k \mathrm{SE}_k(\boldsymbol{\eta})
\quad
\text{s.t. } \eqref{eq:papc}.
\label{eq:maxmin}
\end{equation}
An equivalent target-rate feasibility view is also useful: given $\gamma$, determine whether there exists $\boldsymbol{\eta}$ such that
\begin{equation}
\mathrm{SE}_k(\boldsymbol{\eta}) \ge \gamma, \qquad \forall k,
\label{eq:target}
\end{equation}
under \eqref{eq:papc}. In the prototype, the verifier operates on returned candidate solutions rather than on intermediate solver variables. For a candidate $\hat{\boldsymbol{\eta}}$, the verified common rate is
\begin{equation}
r_{\mathrm{ver}}(\hat{\boldsymbol{\eta}})=\min_k \mathrm{SE}_k(\hat{\boldsymbol{\eta}}),
\label{eq:rver}
\end{equation}
and the candidate is accepted only if the recomputed per-AP power constraints are satisfied and the resulting verified rate or feasibility target satisfies the predefined acceptance criterion.

Let $\mathcal{T}=\{T_1,\ldots,T_M\}$ denote a portfolio of trusted solvers. For each instance, the orchestration layer observes a structured descriptor $\mathbf{x}$ and outputs a solver plan
\begin{equation}
\pi(\mathbf{x})=(t^{(1)},\mathcal{F},b,c),
\label{eq:plan}
\end{equation}
where $t^{(1)}$ is the first solver, $\mathcal{F}$ is an ordered fallback list, $b$ is an optional budget policy, and $c$ denotes acceptance criteria. In the current prototype, $\mathbf{x}$ contains compact indicators of regime difficulty, including problem size, power budget, fairness target, summary statistics of large-scale fading, and user-imbalance descriptors. The verifier recomputes feasibility and verified common rate from the returned candidate solution.

For evaluation, let $v(i,m)$ be the verified common rate obtained by method $m$ on instance $i$, and let $v^\star(i)=\max_m v(i,m)$ denote the ex-post best verified value among compared methods. The instance-level selection regret is
\begin{equation}
\mathrm{regret}(i,m)=v^\star(i)-v(i,m).
\label{eq:regret}
\end{equation}
The orchestration objective is therefore not only to maximize verified success, but also to keep regret, runtime, and fallback cost low across heterogeneous operating conditions. In this correspondence, accepted rate is the fraction of instances for which the final returned solution passes the verifier, whereas fallback rate measures how often the initial solver choice is insufficient and an additional solver call is required. These metrics are central because the contribution is deployment-oriented reliability rather than raw objective value.

\section{Proposed Method}

Given a downlink cell-free power-control instance, the goal is to return a verified solution that satisfies the target constraints and attains strong objective value within a practical runtime budget. The proposed \textbf{VISO-PC} framework therefore treats the agent as a solver orchestrator rather than as a direct power-control policy. The design is inspired by the broader tool-intelligence view of agentic AI in communications~\cite{liu2026unleashing}, but is specialized here to verifier-aware solver routing.

\begin{figure}[t]
    \centering
    \resizebox{0.82\columnwidth}{!}{%
    \begin{tikzpicture}[
        node distance=7mm,
        box/.style={draw, rounded corners, align=center, minimum width=3.0cm, minimum height=0.85cm, fill=blue!5},
        decision/.style={draw, diamond, aspect=2.0, align=center, inner sep=0.5pt, fill=orange!8},
        sidebox/.style={draw, rounded corners, align=center, minimum width=2.5cm, minimum height=0.8cm, fill=green!5},
        line/.style={-latex, thick}
    ]
        \node[box] (inst) {Raw CF-mMIMO instance};
        \node[box, below=of inst] (desc) {Descriptor summarizer};
        \node[box, below=of desc] (router) {Router $\rightarrow$ solver plan};
        \node[box, below=of router] (solver) {Selected solver $t^{(1)}$};
        \node[decision, below=10mm of solver] (check) {Verifier accept?};
        \node[box, below=10mm of check] (final) {Return verified solution};
        \node[sidebox, left=18mm of check] (fallback) {Fallback list $\mathcal{F}$};

        \draw[line] (inst) -- (desc);
        \draw[line] (desc) -- (router);
        \draw[line] (router) -- (solver);
        \draw[line] (solver) -- (check);
        \draw[line] (check) -- node[right, font=\scriptsize]{yes} (final);
        \draw[line] (check) -- node[above, font=\scriptsize]{no} (fallback);
        \draw[line] (fallback) |- node[pos=0.25, above, font=\scriptsize]{retry} (solver);
    \end{tikzpicture}%
    }
    \caption{Overview of VISO-PC. A raw power-control instance is summarized into a structured descriptor, routed to a solver portfolio, and checked by an independent verifier. If the current solver fails verification, the fallback plan is invoked until a verified solution is found or the budget is exhausted.}
    \label{fig:pipeline}
\end{figure}
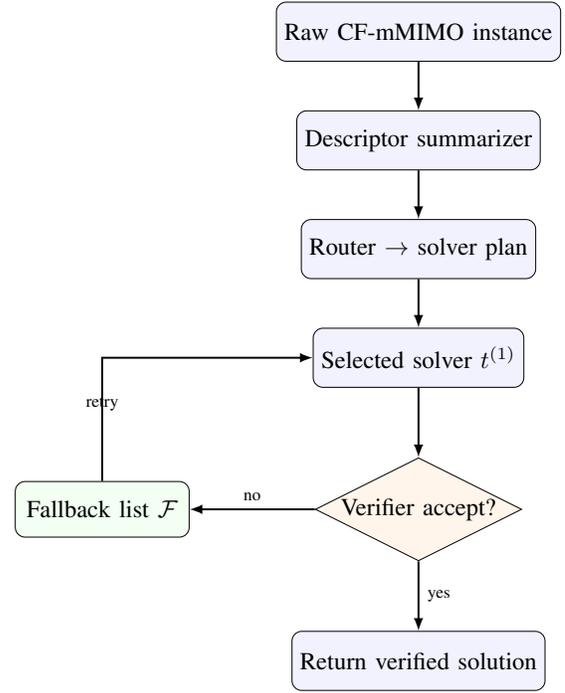

The solver portfolio contains three complementary solvers. The first, $T_{\text{fast}}$, is a low-complexity iterative fairness solver that serves as the default low-latency option. The second, $T_{\text{exact}}$, is a slower but stronger feasibility-oriented solver surrogate. The third, $T_{\text{dist}}$, is a low-cost distributed-style heuristic used mainly as a fallback. This design follows the classical observation that cell-free implementations exhibit distinct quality-scalability trade-offs~\cite{Nayebi2017PrecodingPowerOptimization,Chakraborty2019CentralizedDistributedMaxMin,Bjornson2020ScalableCellFree}.

The router maps the instance descriptor to a solver plan as in \eqref{eq:plan}. The descriptor contains compact features such as problem size, fairness target, power budget, large-scale fading statistics, and user-imbalance indicators. In the current prototype, Rule-Router uses hand-designed thresholds, whereas Agent-Router performs nearest-neighbor retrieval over an oracle-labeled memory built from the train split. Each memory entry stores a descriptor together with the verified best solver on the corresponding training instance. At inference time, retrieval determines the initial solver preference, after which the orchestrator follows the ordered fallback list if the first attempt is rejected. This keeps the agentic component lightweight and interpretable while still enabling instance-dependent routing.

For solver $j$, let $\hat{\boldsymbol{\eta}}^{(j)}$ denote the returned candidate allocation and let $a^{(j)}\in\{0,1\}$ denote whether the verifier accepts it. In the current prototype, the budget policy $b$ is implemented as a sequential-call budget over the ordered solver list: the orchestrator evaluates one solver at a time and stops when a candidate is accepted or the list is exhausted. The acceptance rule $c$ is verifier-based. A returned candidate is accepted only if the independent checker recomputes the per-AP constraints in \eqref{eq:papc} and confirms a valid verified common rate. The orchestration decision is therefore governed by
\begin{equation}
a^{(j)} = \mathbf{1}\!\left[\hat{\boldsymbol{\eta}}^{(j)}\text{ satisfies }\eqref{eq:papc}\text{ and }c\right],
\label{eq:accept}
\end{equation}
where $c$ denotes the verifier-based acceptance rule described above. If the current solver returns no usable candidate, fails verification, or exhausts the remaining budget before acceptance, the orchestrator invokes the next solver in the fallback order. Formally, if $\hat{\boldsymbol{\eta}}^{(j)}$ is the candidate from the $j$-th attempted solver, the final returned solution is
\begin{equation}
\hat{\boldsymbol{\eta}}^{\mathrm{out}} = \hat{\boldsymbol{\eta}}^{(j^\star)},
\qquad
j^\star = \min\{j : a^{(j)}=1\},
\label{eq:earlystop}
\end{equation}
if such an index exists; otherwise the instance is declared unresolved after the fallback chain is exhausted. Thus, fallback is not a parallel voting mechanism: solvers are tried in sequence, and the first candidate that passes independent verification is returned immediately. Early stopping is therefore a core part of the method rather than an implementation detail.

The practical orchestration objective can be expressed as selecting a plan that balances verified quality and runtime,
\begin{equation}
\max_{\pi(\mathbf{x})} \; \mathbb{E}\!\left[r_{\mathrm{ver}}(\hat{\boldsymbol{\eta}}^{\mathrm{out}}) - \lambda \tau(\pi(\mathbf{x}))\right],
\label{eq:tradeoff}
\end{equation}
where $\tau(\pi(\mathbf{x}))$ is the total runtime induced by the chosen fallback path and $\lambda$ is a trade-off coefficient. Although the prototype does not optimize \eqref{eq:tradeoff} directly, this expression motivates the evaluation metrics.

This design separates two error sources that might otherwise be confounded: weak routing and weak solver capability. As a result, the evaluation can reveal whether a difficult regime is router-limited or solver-limited. It also makes total runtime approximately additive over attempted solver calls, so fallback frequency and average attempts become operational deployment metrics rather than bookkeeping details.

\subsection{Router Instantiation and Complexity}
The Rule-Router is a descriptor-based heuristic that applies hand-crafted thresholds to prioritize either the fast solver or the stronger exact solver, while keeping a fixed fallback order. The Agent-Router replaces these thresholds with nearest-neighbor retrieval over the train-split memory. The agentic component in the current prototype is intentionally lightweight: it does not synthesize new optimization logic, but instead retrieves a solver preference from prior verified experience.

The online orchestration overhead is modest. Descriptor extraction is linear in the number of summary features, routing is negligible relative to solver runtime, and verification is dominated by recomputing the target metrics of the returned candidate. Because the prototype returns the first accepted candidate, the realized path is determined by the verifier outcome and the fallback order. Consequently, the main practical cost is fallback depth. This is why reducing the average number of attempts is valuable even when accepted rate remains unchanged. From a deployment perspective, the architecture is attractive because it preserves the trustworthiness of classical solvers while allowing an agentic layer to adapt solver usage to heterogeneous instances.

\begin{algorithm}[t]
\caption{Agentic Verifier-in-the-Loop Solver Orchestration}
\label{alg:viso}
\footnotesize
\begin{algorithmic}[1]
\State Compute descriptor $\mathbf{x}$ from instance $\mathcal{I}$
\State Obtain solver plan $\pi(\mathbf{x})=(t^{(1)},\mathcal{F},b,c)$
\State Form ordered solver list $[t^{(1)};\mathcal{F}]$
\State Initialize cumulative runtime $\tau \leftarrow 0$
\For{each solver $t$ in the ordered list}
    \State Run solver $t$ and obtain candidate $\hat{\boldsymbol{\eta}}^{(j)}$ and runtime increment $\Delta \tau$
    \State Update cumulative runtime $\tau \leftarrow \tau + \Delta \tau$
    \State Verify feasibility and compute $r_{\mathrm{ver}}(\hat{\boldsymbol{\eta}}^{(j)})$
    \State Compute acceptance flag $a^{(j)}$
    \If{$a^{(j)}=1$}
        \State \Return $\hat{\boldsymbol{\eta}}^{(j)}$
    \EndIf
    \If{runtime budget is exhausted}
        \State \textbf{break}
    \EndIf
\EndFor
\State Return unresolved status
\end{algorithmic}
\end{algorithm}

\section{Simulation Results and Analysis}

\subsection{Simulation Setup}
We evaluate VISO-PC on a reproducible benchmark of downlink cell-free massive MIMO power-control instances. The prototype uses a restricted but representative setting with single-antenna APs and UEs, fixed maximum-ratio transmission, and per-AP power constraints. All methods are evaluated on the same instance pool and checked by the same independent verifier, so any performance difference reflects solver behavior and routing quality rather than inconsistent post-processing.

Instances are organized into four regimes with distinct roles. The train split is used only to build the Agent-Router memory. The test split evaluates in-distribution generalization within the same benchmark family. The stress split contains harder instances that expose solver limitations even when routing is reasonable. The shifted split alters the descriptor-to-best-solver relationship and therefore probes retrieval robustness under distribution shift. This split design allows the later analysis to distinguish router-limited behavior from solver-limited behavior.

The portfolio consists of $T_{\text{fast}}$, $T_{\text{exact}}$, and $T_{\text{dist}}$. Always-Fast, Always-Exact, and Always-Dist commit to one solver for every instance. Rule-Router selects the first solver by hand-designed descriptor thresholds and then follows a fixed fallback order. Agent-Router uses nearest-neighbor retrieval over the train-split memory, whose entries are oracle-labeled by the verified best solver choice on the corresponding training instances. The retrieved preference determines the first solver, and the remaining solvers define the fallback path.

The reported metrics are accepted rate, feasible rate, verified common rate, runtime, selection regret, fallback rate, and average attempts. Accepted rate is the fraction of instances for which the final returned candidate passes the verifier. Feasible rate reports the fraction of instances whose returned candidate is verifier-feasible, and in the present benchmark it coincides numerically with accepted rate because acceptance is verifier-defined. Verified common rate is always recomputed from the returned candidate rather than taken from the raw solver output. The benchmark is deliberately small-scale because the goal of this correspondence is not to claim a final large-scale performance record, but to present a clear and reproducible verifier-aware orchestration formulation for communication systems.

\subsection{Overall Results}
Table~\ref{tab:main_results} reports the overall comparison. Both routing methods outperform the fixed single-solver policies in accepted rate. Rule-Router and Agent-Router each achieve an accepted rate of 0.6923, whereas Always-Fast reaches 0.6154 and Always-Exact reaches 0.5385. In this benchmark, feasible rate equals accepted rate because acceptance is verifier-based. Verification-aware orchestration is therefore more reliable than committing to a single solver across all instances.

\begin{figure}[t]
    \centering
    \includegraphics[width=\columnwidth]{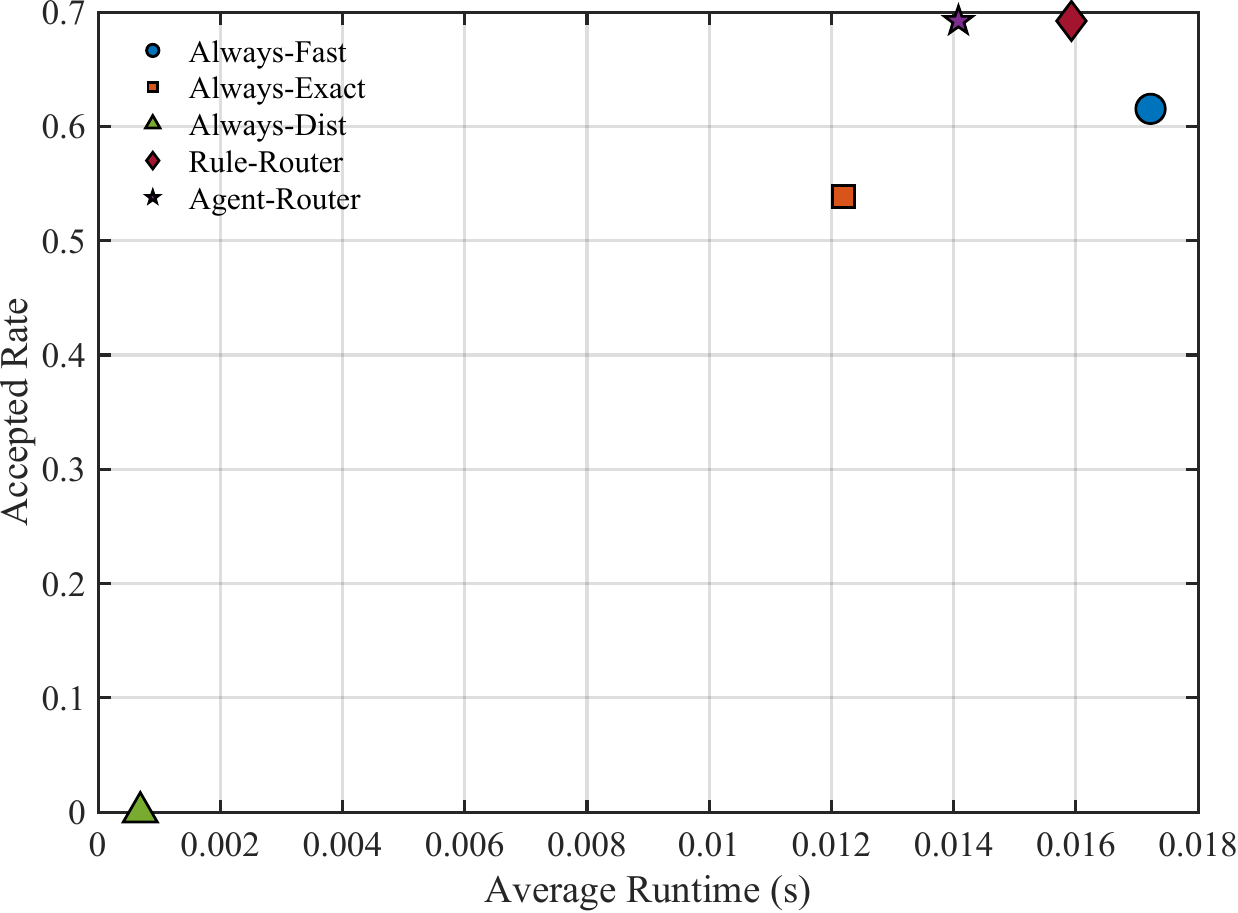}
    \caption{Accepted-rate versus average-runtime trade-off on the prototype benchmark. Router-based orchestration improves accepted rate over all fixed single-solver policies. Agent-Router matches the accepted rate of Rule-Router while using lower average runtime, indicating a favorable efficiency trade-off after early stopping is enabled.}
    \label{fig:main_frontier}
\end{figure}

Rule-Router remains the only zero-regret method, but Agent-Router matches its accepted rate while reducing average runtime, fallback rate, and average attempts. Fig.~\ref{fig:main_frontier} shows this overall frontier. Table~\ref{tab:main_results} also clarifies why orchestration is necessary: Always-Exact is strong in solution quality but does not maximize accepted rate, whereas Always-Fast solves more cases but incurs larger regret. Always-Dist is the cheapest method in runtime but offers little verified performance on the current benchmark, which justifies its role as a low-cost fallback rather than a primary solver.

\begin{table*}[t]
\centering
\caption{Overall comparison on the prototype benchmark. Higher is better for accepted rate, feasible rate, and verified common rate, while lower is better for average runtime, average regret, fallback rate, and average attempts.}
\label{tab:main_results}
\begin{tabular}{lccccccc}
\toprule
Method & Acc. Rate & Feas. Rate & Ver. Common Rate & Avg. Runtime & Avg. Regret & Fallback Rate & Avg. Attempts \\
\midrule
Always-Fast  & 0.6154 & 0.6154 & 2.48e-05 & 0.01722 & 4.64e-05 & 0.0000 & 1.0000 \\
Always-Exact & 0.5385 & 0.5385 & 4.96e-05 & 0.01220 & 7.59e-06 & 0.0000 & 1.0000 \\
Always-Dist  & 0.0000 & 0.0000 & 3.63e-08 & \textbf{0.00069} & 7.79e-05 & 0.0000 & 1.0000 \\
Rule-Router  & \textbf{0.6923} & \textbf{0.6923} & \textbf{7.80e-05} & 0.01593 & \textbf{0.0000} & 0.4615 & 1.7692 \\
\textbf{Agent-Router} & \textbf{0.6923} & \textbf{0.6923} & 6.96e-05 & \textbf{0.01408} & 8.42e-06 & \textbf{0.3077} & \textbf{1.6154} \\
\bottomrule
\end{tabular}
\end{table*}

\subsection{Split-wise and Failure Analysis}
\begin{figure}[t]
    \centering
    \includegraphics[width=\columnwidth]{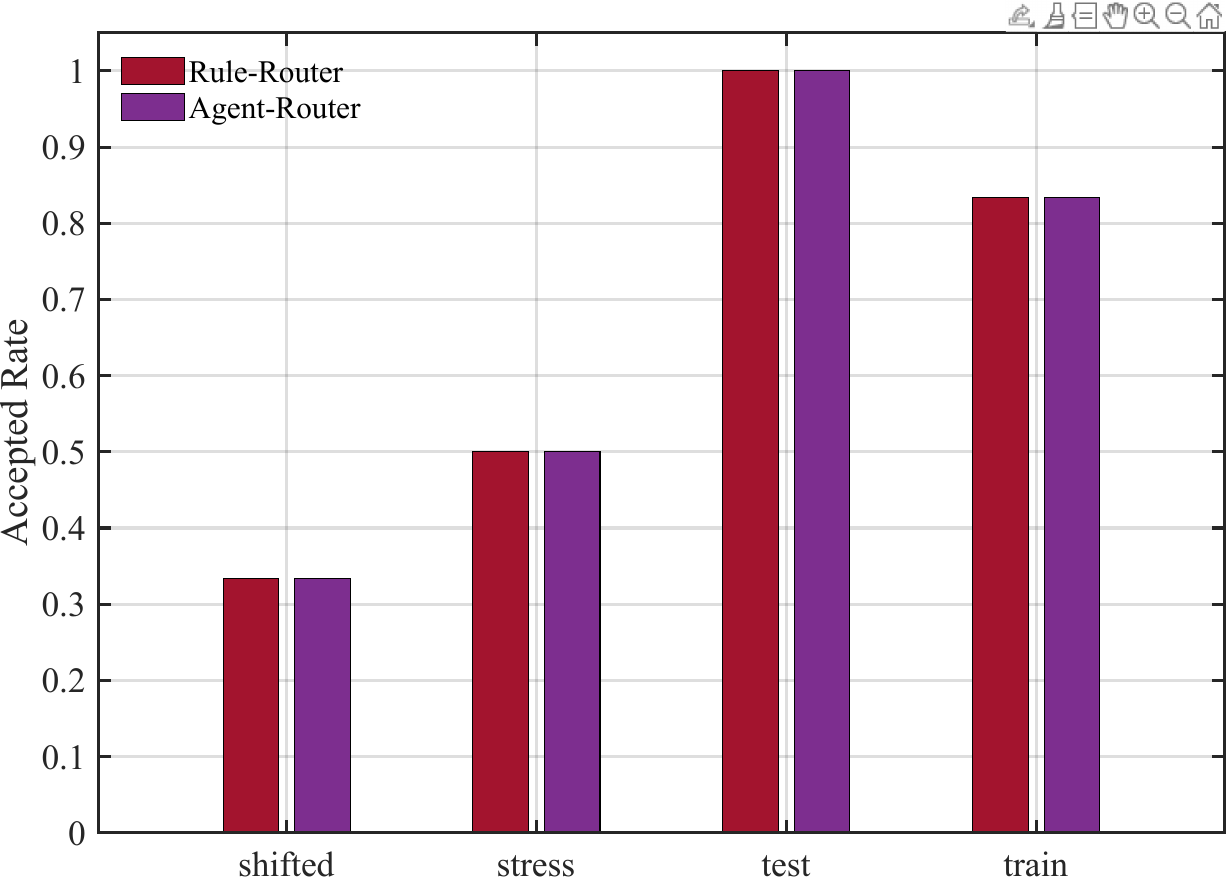}
    \caption{Split-wise accepted-rate comparison for the two routing methods. The accepted rates match on all four splits, so the main differences between the routers appear in runtime and regret rather than in acceptance itself. This pattern is consistent with stress being solver-limited and shifted cases exposing retrieval fragility.}
    \label{fig:split_bars}
\end{figure}

Table~\ref{tab:split_results} reports split-wise results. On train and test, Agent-Router matches the accepted rate of Rule-Router while using lower runtime and fewer fallback steps, indicating that the current descriptor-memory matching is already effective in in-distribution regimes. On stress, both routers behave similarly, indicating that these cases are solver-limited rather than router-limited. On shifted, the two routers again match in accepted rate, but Agent-Router incurs higher regret, indicating that retrieval remains fragile under distribution shift. Fig.~\ref{fig:split_bars} visualizes this regime dependence.

\begin{table*}[t]
\centering
\caption{Split-wise comparison of fixed solvers and router-based methods. Higher is better for accepted rate and verified common rate, while lower is better for average runtime, average regret, fallback rate, and average attempts.}
\label{tab:split_results}
\begin{tabular}{llcccccc}
\toprule
Split & Method & Acc. Rate & Ver. Common Rate & Avg. Runtime & Avg. Regret & Fallback Rate & Avg. Attempts \\
\midrule
Train & Always-Fast  & 0.6667 & 2.32e-05 & 0.01512 & 4.40e-05 & 0.0000 & 1.0000 \\
Train & Always-Exact & 0.5000 & 4.79e-05 & 0.00964 & 1.40e-05 & 0.0000 & 1.0000 \\
Train & Rule-Router  & \textbf{0.8333} & \textbf{7.06e-05} & 0.01314 & \textbf{0.0000} & 0.5000 & 1.6667 \\
Train & \textbf{Agent-Router} & \textbf{0.8333} & \textbf{7.06e-05} & \textbf{0.00964} & \textbf{0.0000} & \textbf{0.1667} & \textbf{1.3333} \\
\midrule
Test & Always-Fast  & 0.8750 & 3.88e-05 & 0.01546 & 5.51e-05 & 0.0000 & 1.0000 \\
Test & Always-Exact & 0.8750 & 8.79e-05 & 0.00985 & 6.01e-06 & 0.0000 & 1.0000 \\
Test & Rule-Router  & \textbf{1.0000} & \textbf{9.39e-05} & 0.01111 & \textbf{0.0000} & 0.1250 & 1.1250 \\
Test & \textbf{Agent-Router} & \textbf{1.0000} & 8.27e-05 & \textbf{0.00860} & 1.12e-05 & \textbf{0.0000} & \textbf{1.0000} \\
\midrule
Stress & Always-Fast  & 0.5000 & 2.13e-05 & 0.02078 & 3.72e-05 & 0.0000 & 1.0000 \\
Stress & Always-Exact & 0.5000 & 3.31e-05 & \textbf{0.01562} & \textbf{0.0000} & 0.0000 & 1.0000 \\
Stress & Rule-Router  & \textbf{0.5000} & \textbf{6.61e-05} & 0.02028 & \textbf{0.0000} & 0.5000 & 2.0000 \\
Stress & \textbf{Agent-Router} & \textbf{0.5000} & \textbf{6.61e-05} & 0.02193 & \textbf{0.0000} & 0.5000 & 2.0000 \\
\midrule
Shifted & Always-Fast  & 0.3333 & 1.13e-05 & 0.01812 & 3.09e-05 & 0.0000 & 1.0000 \\
Shifted & Always-Exact & 0.1667 & 1.66e-05 & \textbf{0.01446} & 9.18e-06 & 0.0000 & 1.0000 \\
Shifted & Rule-Router  & \textbf{0.3333} & \textbf{5.06e-05} & 0.02079 & \textbf{0.0000} & 0.8333 & 2.5000 \\
Shifted & \textbf{Agent-Router} & \textbf{0.3333} & 1.96e-05 & \textbf{0.01796} & 3.09e-05 & \textbf{0.6667} & \textbf{2.3333} \\
\bottomrule
\end{tabular}
\end{table*}

Table~\ref{tab:id_ood} aggregates the split-wise results into in-distribution (Train+Test) and out-of-distribution (Stress+Shifted) regimes to clarify the regime dependence. For in-distribution cases, Agent-Router matches the accepted rate of Rule-Router while reducing runtime and fallback rate, showing that lightweight memory retrieval is already effective on familiar instances. For out-of-distribution cases, the accepted rate remains matched, but Rule-Router retains a higher verified common rate and zero regret. The trade-off is therefore clear: the current agentic layer improves efficiency, whereas robustness under shift still depends on stronger retrieval and solver diversity.

\begin{table}[t]
\centering
\caption{Aggregated in-distribution (ID) and out-of-distribution (OOD) comparison for the two routing methods. ID aggregates Train and Test, and OOD aggregates Stress and Shifted. Higher is better for accepted rate and verified common rate, while lower is better for average runtime, average regret, and fallback rate.}
\label{tab:id_ood}
\resizebox{\columnwidth}{!}{%
\begin{tabular}{lcccccc}
\toprule
Group & Method & Acc. Rate & Ver. Common Rate & Avg. Runtime & Avg. Regret & Fallback Rate \\
\midrule
ID  & Rule  & 0.9286 & \textbf{8.39e-05} & 0.01198 & 0.0000 & 0.2857 \\
ID  & \textbf{Agent} & 0.9286 & 7.75e-05 & \textbf{0.00905} & 6.41e-06 & \textbf{0.0714} \\
OOD & Rule  & 0.4167 & \textbf{5.83e-05} & 0.02053 & 0.0000 & 0.6667 \\
OOD & \textbf{Agent} & 0.4167 & 4.29e-05 & \textbf{0.01995} & 1.55e-05 & \textbf{0.5833} \\
\bottomrule
\end{tabular}%
}
\end{table}

\begin{figure}[t]
    \centering
    \includegraphics[width=\columnwidth]{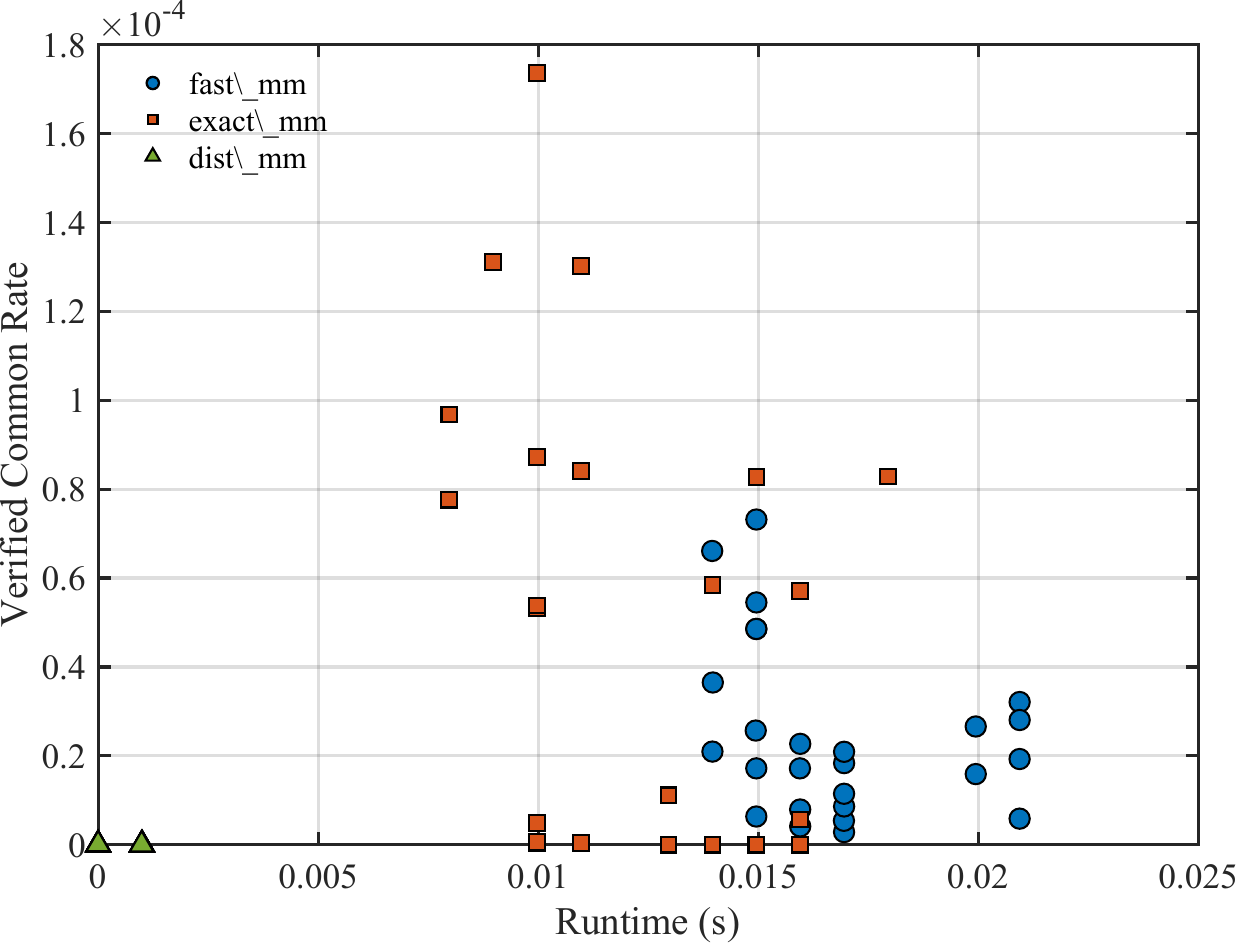}
    \caption{Per-instance solver profiling in average-runtime--verified-common-rate space. Exact-solver instances occupy the strongest quality region, fast-solver instances cover many easier cases at lower runtime, and the distributed solver remains a low-cost but weak fallback. The spread across solvers motivates orchestration over a solver portfolio.}
    \label{fig:solver_profile}
\end{figure}

Fig.~\ref{fig:solver_profile} further shows that the portfolio is genuinely complementary. The exact solver occupies the strongest quality region, the fast solver covers many easier instances at lower runtime, and the distributed solver remains a low-cost but weak fallback. Orchestration is meaningful because the available solvers populate different quality-latency regions. The proposed agentic layer does not replace optimization theory; it exploits the operating-region diversity of existing optimization tools.

Table~\ref{tab:failure_cases} lists representative shifted failure cases. On \texttt{inst\_shifted\_303}, Rule-Router selects \texttt{exact\_mm} and attains a much higher verified common rate, whereas Agent-Router selects \texttt{fast\_mm} and solves the instance more quickly but with a larger gap. This behavior matches the split-wise regret pattern and suggests that future gains are more likely to come from stronger retrieval and solver diversity than from changing the orchestration loop itself. Taken together, the results indicate a clear division of labor: verification guarantees correctness, routing improves efficiency in familiar regimes, and stronger solvers remain necessary in shifted or otherwise difficult operating regions.

\begin{table}[t]
\centering
\caption{Representative shifted failure cases. Agent-Router can reduce runtime on some out-of-distribution instances, but Rule-Router can obtain a higher verified common rate when it selects the stronger solver earlier.}
\label{tab:failure_cases}
\resizebox{\columnwidth}{!}{%
\begin{tabular}{lccccc}
\toprule
Instance & Method & Solver & Accepted & Ver. Common Rate & Runtime \\
\midrule
301 & Rule  & fast  & T & 1.84e-05 & 0.02197 \\
301 & \textbf{Agent} & fast  & T & 1.84e-05 & \textbf{0.00895} \\
303 & Rule  & exact & T & \textbf{8.27e-05} & 0.01396 \\
303 & \textbf{Agent} & fast  & T & 2.09e-05 & \textbf{0.00900} \\
300 & Rule  & --    & F & -- & 0.02195 \\
300 & \textbf{Agent} & --    & F & -- & 0.02297 \\
\bottomrule
\end{tabular}%
}
\end{table}

\section{Conclusion}

This paper presented VISO-PC, a verifier-in-the-loop solver-orchestration framework for cell-free massive MIMO downlink power control. The key idea is to use the agent as a routing layer over trusted solvers, with independent verification and fallback, rather than to generate power coefficients directly. On the prototype benchmark, orchestration improves accepted rate over fixed single-solver baselines, and the lightweight memory-based router matches the accepted rate of a strong rule-based router while reducing average runtime and fallback rate. The results also reveal a clear regime split: stress cases are primarily solver-limited, whereas shifted cases are retrieval-limited. Although the current study is intentionally narrow---downlink, fairness-oriented power control with fixed beamforming---it establishes solver orchestration as a reproducible formulation for communication systems. Future work can expand the solver portfolio, strengthen retrieval under distribution shift, and extend the framework to richer objectives and larger cell-free settings.

\bibliographystyle{IEEEtran}
\bibliography{references}

@inproceedings{Ngo2015CellFreeUniformlyGreat,
  title={Cell-Free Massive {MIMO}: Uniformly Great Service for Everyone},
  author={Ngo, Hien Quoc and Ashikhmin, Alexei and Yang, Hong and Larsson, Erik G. and Marzetta, Thomas L.},
  booktitle={2015 IEEE 16th International Workshop on Signal Processing Advances in Wireless Communications (SPAWC)},
  pages={201--205},
  year={2015}
}

@article{Ngo2017CellFreeVersusSmallCells,
  title={Cell-Free Massive {MIMO} Versus Small Cells},
  author={Ngo, Hien Quoc and Ashikhmin, Alexei and Yang, Hong and Larsson, Erik G. and Marzetta, Thomas L.},
  journal={IEEE Transactions on Wireless Communications},
  volume={16},
  number={3},
  pages={1834--1850},
  year={2017}
}

@article{Nayebi2017PrecodingPowerOptimization,
  title={Precoding and Power Optimization in Cell-Free Massive {MIMO} Systems},
  author={Nayebi, Elina and Ashikhmin, Alexei and Marzetta, Thomas L. and Yang, Hong and Rao, Bhaskar D.},
  journal={IEEE Transactions on Wireless Communications},
  volume={16},
  number={7},
  pages={4445--4459},
  year={2017}
}

@inproceedings{Chakraborty2019CentralizedDistributedMaxMin,
  title={Centralized and Distributed Power Allocation for Max-Min Fairness in Cell-Free Massive {MIMO}},
  author={Chakraborty, Sucharita and Bjornson, Emil and Sanguinetti, Luca},
  booktitle={2019 53rd Asilomar Conference on Signals, Systems, and Computers},
  pages={576--580},
  year={2019}
}

@article{Bjornson2020ScalableCellFree,
  title={Scalable Cell-Free Massive {MIMO} Systems},
  author={Bjornson, Emil and Sanguinetti, Luca},
  journal={IEEE Transactions on Communications},
  volume={68},
  number={7},
  pages={4247--4261},
  year={2020}
}

@article{Bjornson2020MakingCellFreeCompetitive,
  title={Making Cell-Free Massive {MIMO} Competitive With {MMSE} Processing and Centralized Implementation},
  author={Bjornson, Emil and Sanguinetti, Luca},
  journal={IEEE Transactions on Wireless Communications},
  volume={19},
  number={1},
  pages={77--90},
  year={2020}
}

@misc{shokouhi2026agentic,
  title={Agentic {AI} for Intent-driven Optimization in Cell-free {O-RAN}},
  author={Mohammad Hossein Shokouhi and Vincent W. S. Wong},
  year={2026},
  eprint={2602.22539},
  archivePrefix={arXiv},
  primaryClass={cs.AI}
}

@misc{liu2026unleashing,
  title={Unleashing Tool Engineering and Intelligence for Agentic {AI} in Next-Generation Communication Networks},
  author={Yinqiu Liu and Ruichen Zhang and Dusit Niyato and Abbas Jamalipour and Trung Q. Duong and Dong In Kim},
  year={2026},
  eprint={2601.08259},
  archivePrefix={arXiv},
  primaryClass={cs.NI}
}

@misc{elkael2025agentran,
  title={AgentRAN: An Agentic {AI} Architecture for Autonomous Control of Open {6G} Networks},
  author={Maxime Elkael and Salvatore D'Oro and Leonardo Bonati and Michele Polese and Yunseong Lee and Koichiro Furueda and Tommaso Melodia},
  year={2025},
  eprint={2508.17778},
  archivePrefix={arXiv},
  primaryClass={cs.AI}
}

@misc{elkael2025allstar,
  title={{ALLSTaR}: Automated {LLM}-Driven Scheduler Generation and Testing for Intent-Based {RAN}},
  author={Maxime Elkael and Michele Polese and Reshma Prasad and Stefano Maxenti and Tommaso Melodia},
  year={2025},
  eprint={2505.18389},
  archivePrefix={arXiv},
  primaryClass={cs.NI}
}

@misc{navidan2026toward,
  title={Toward Autonomous {O-RAN}: A Multi-Scale Agentic {AI} Framework for Real-Time Network Control and Management},
  author={Hojjat Navidan and Mohammad Cheraghinia and Jaron Fontaine and Mohamed Seif and Eli De Poorter and H. Vincent Poor and Ingrid Moerman and Adnan Shahid},
  year={2026},
  eprint={2602.14117},
  archivePrefix={arXiv},
  primaryClass={cs.NI}
}

@misc{li2026multiagentic,
  title={Multi-Agentic {AI} for Conflict-Aware rApp Policy Orchestration in Open {RAN}},
  author={Haiyuan Li and Yulei Wu and Dimitra Simeonidou},
  year={2026},
  eprint={2603.07375},
  archivePrefix={arXiv},
  primaryClass={eess.SY}
}

@misc{peng2025llmoptira,
  title={{LLM}-{OptiRA}: {LLM}-Driven Optimization of Resource Allocation for Non-Convex Problems in Wireless Communications},
  author={Xinyue Peng and Yanming Liu and Yihan Cang and Chaoqun Cao and Ming Chen},
  year={2025},
  eprint={2505.02091},
  archivePrefix={arXiv},
  primaryClass={cs.CL}
}

@misc{zhou2024llmenabled,
  title={Large Language Model ({LLM})-enabled In-context Learning for Wireless Network Optimization: A Case Study of Power Control},
  author={Hao Zhou and Chengming Hu and Dun Yuan and Ye Yuan and Di Wu and Xue Liu and Jianzhong Zhang},
  year={2024},
  eprint={2408.00214},
  archivePrefix={arXiv},
  primaryClass={eess.SY}
}

\end{document}